\documentclass[twocolumn]{article}
\usepackage[utf8]{inputenc}
\usepackage{graphicx}
\usepackage{parskip}
\usepackage{xspace}
\usepackage[tmargin=1in,bmargin=1in,lmargin=1in,
rmargin=1in]{geometry}
\usepackage{hyperref}
\usepackage[sort]{cite}

\title{Teaching Computer Vision for Ecology}
\author{
Elijah Cole$^1$ \qquad Suzanne Stathatos$^1$ \qquad Bj\"orn L\"utjens$^2$ \qquad Tarun Sharma$^1$\\
Justin Kay$^{1,3}$ \qquad Jason Parham$^4$ \qquad Benjamin Kellenberger$^5$ \qquad Sara Beery$^{1,2}$\\
{\small $^1$Caltech \qquad $^2$MIT \qquad $^3$Ai.Fish \qquad $^4$Wild Me \qquad $^5$Yale University}\\
\url{https://cv4ecology.caltech.edu/}
}
\date{}

\newcommand*{\ie}{i.e.\@\xspace}
\newcommand*{\eg}{e.g.\@\xspace}

\begin{document}

\maketitle

\abstract{
Computer vision can accelerate ecology research by automating the analysis of raw imagery from sensors like camera traps, drones, and satellites. However, computer vision is an emerging discipline that is rarely taught to ecologists. This work discusses our experience teaching a diverse group of ecologists to prototype and evaluate computer vision systems in the context of an intensive hands-on summer workshop. We explain the workshop structure, discuss common challenges, and propose best practices. This document is intended for computer scientists who teach computer vision across disciplines, but it may also be useful to ecologists or other domain experts who are learning to use computer vision themselves. 
}

\section{Introduction}

Extracting important information from images and videos normally requires painstaking manual effort from human annotators. Computer vision algorithms can automate this process. This is especially important when manually reviewing the data is not feasible, either because the amount of data is too large (\eg the $>$100TB of satellite imagery collected daily) or the number of annotators is too small (\eg when expertise is required to identify a species in an image). Both of these challenges are common in ecology.

Ecology presents a particularly compelling use case for computer vision. Due to the effects of climate change, we need to monitor animal populations, vegetation properties, and other indicators of ecosystem health at a large scale~\cite{tuia2022perspectives}. Ecologists are collecting vast amounts of raw data with camera traps, drones, and satellites, but there are not enough experts to annotate the data. Computer vision algorithms can accelerate the pace of research in ecology by efficiently transforming this raw data into useful knowledge. Encouraging progress is already being made in areas like animal detection~\cite{beery2019efficient,parham2018detection,kellenberger2018detecting}, fine-grained species recognition~\cite{van2018inaturalist}, individual re-identification~\cite{stewart2021curation}, species distribution modeling~\cite{cole2020geolifeclef, deneu2022very}, and land cover mapping~\cite{robinson2019large}. These efforts can be viewed in the broader context of computational sustainability~\cite{gomes2019computational} and efforts to use machine learning to combat the effects of climate change~\cite{rolnick2022tackling}. 

\begin{figure}[tb]
\centering
\includegraphics[width=0.49\textwidth]{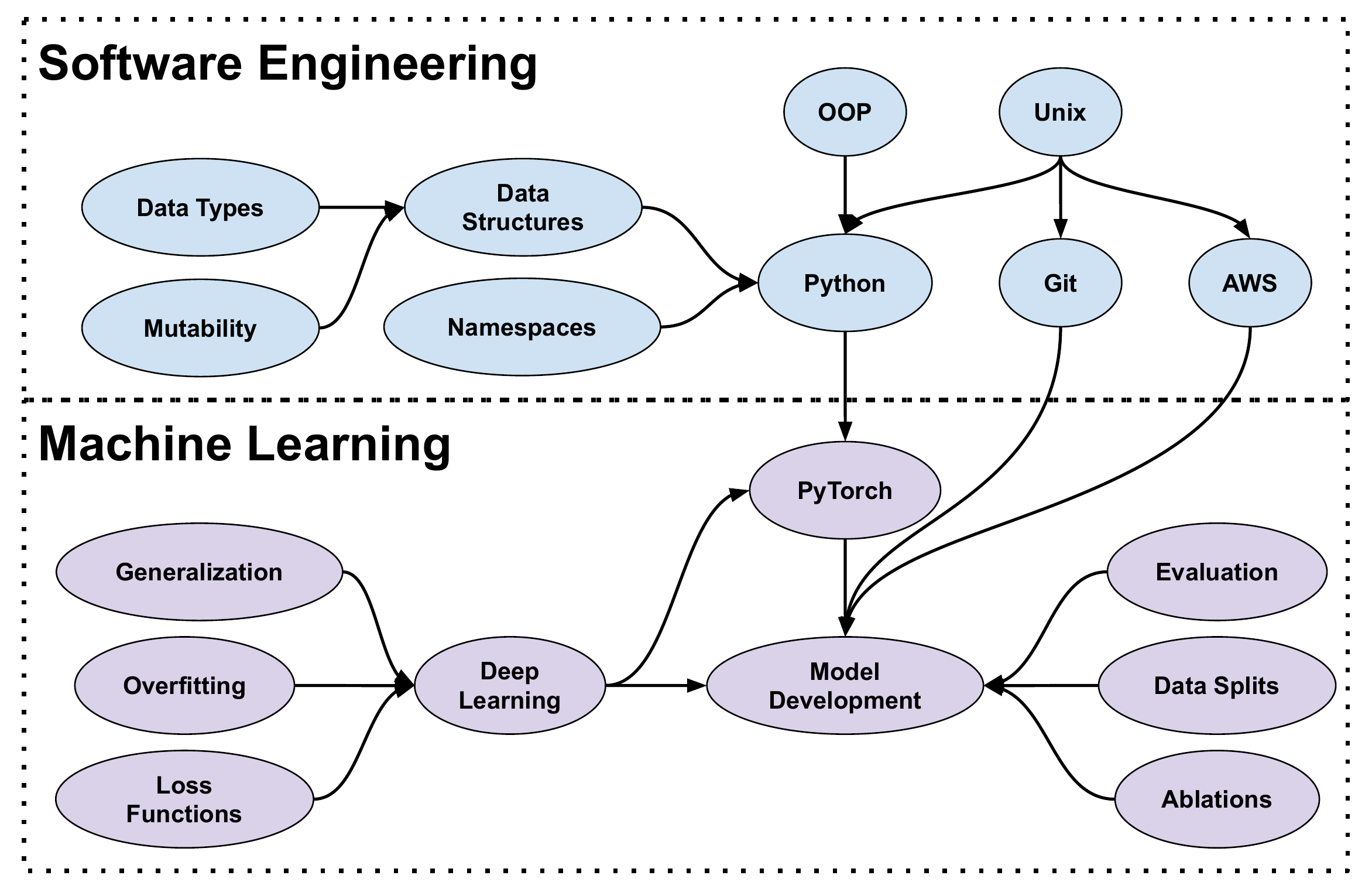}
\caption{A simplified dependency graph depicting some of the skills required to develop a computer vision system. These software engineering and machine learning topics are rarely included in ecology training. See Appendix~\ref{appendix:key_skills} for a catalog of key topics and their significance.}
\label{fig:topics}
\end{figure}

To build on this progress, we must equip ecologists with the skills they need to understand and apply computer vision methods in their research. While ecologists often have training in statistics and programming, they are rarely exposed to the interconnected web of software engineering and machine learning topics necessary for computer vision. We illustrate a few of these topics in Figure~\ref{fig:topics}. 

In this work, we discuss the process of teaching computer vision to ecologists in the context of the \emph{Resnick Sustainability Institute Summer Workshop on Computer Vision Methods for Ecology} (CV4E Workshop), an intensive 3-week workshop held at Caltech in 2022~\cite{cv4e_web}. We review related work in Section~\ref{sec:related_work} before describing the workshop in Section~\ref{sec:workshop_description}, discussing key take-aways in Section~\ref{sec:lessons_learned}, and outlining educational techniques we found useful in Section~\ref{sec:educational_techniques}. 

\section{Related Work}\label{sec:related_work}

There is an emerging literature devoted to teaching machine learning~\cite{steinbach2021teaching, de2021m, shouman2022experiences}, deep learning~\cite{prince2022deeplearning}, and computer vision~\cite{prince2012cvmodels, hassner2015teaching, khorbotly2015project, spurlock2017making}. \cite{skripchuk2022identifying} more narrowly focuses on common errors in machine learning course projects. However, most of these works concern efforts to teach students from computer science or related disciplines. There is prior work discussing the specific challenge of teaching machine learning to cross-disciplinary audiences such as non-CS undergraduates~\cite{sulmont2019hard}, business students~\cite{wunderlich2021machine}, artists~\cite{fiebrink2019machine}, materials scientists~\cite{sun2022teaching}, and biologists~\cite{magnano2022approachable}. Our work is complementary, focusing on the process of teaching computer vision to ecologists (mostly Ph.D. students and postdocs -- see Figure~\ref{fig:cohort}) who have background knowledge in statistics and programming but little prior experience in machine learning. In addition, we consider an immersive workshop in which researchers build prototypes using their own research data, not a traditional classroom environment. 

\section{The CV4E Workshop}\label{sec:workshop_description}

The inaugural CV4E Workshop was held at Caltech from August 1~-~19, 2022. The program was designed to train ecologists to use computer vision in their own research. Here we outline the stages of the workshop.

\textbf{Application.} The application had five components: (i) a one-page project proposal, (ii) a one-page personal statement, (iii) a programming example, (iv) one letter of reference, and (v) a CV. The most important element was the project proposal, in which participants described the problem they wanted to solve with computer vision, the potential impact of a working solution, and the available data and labels. 

\begin{figure}[htb]
\centering
\includegraphics[width=0.49\textwidth]{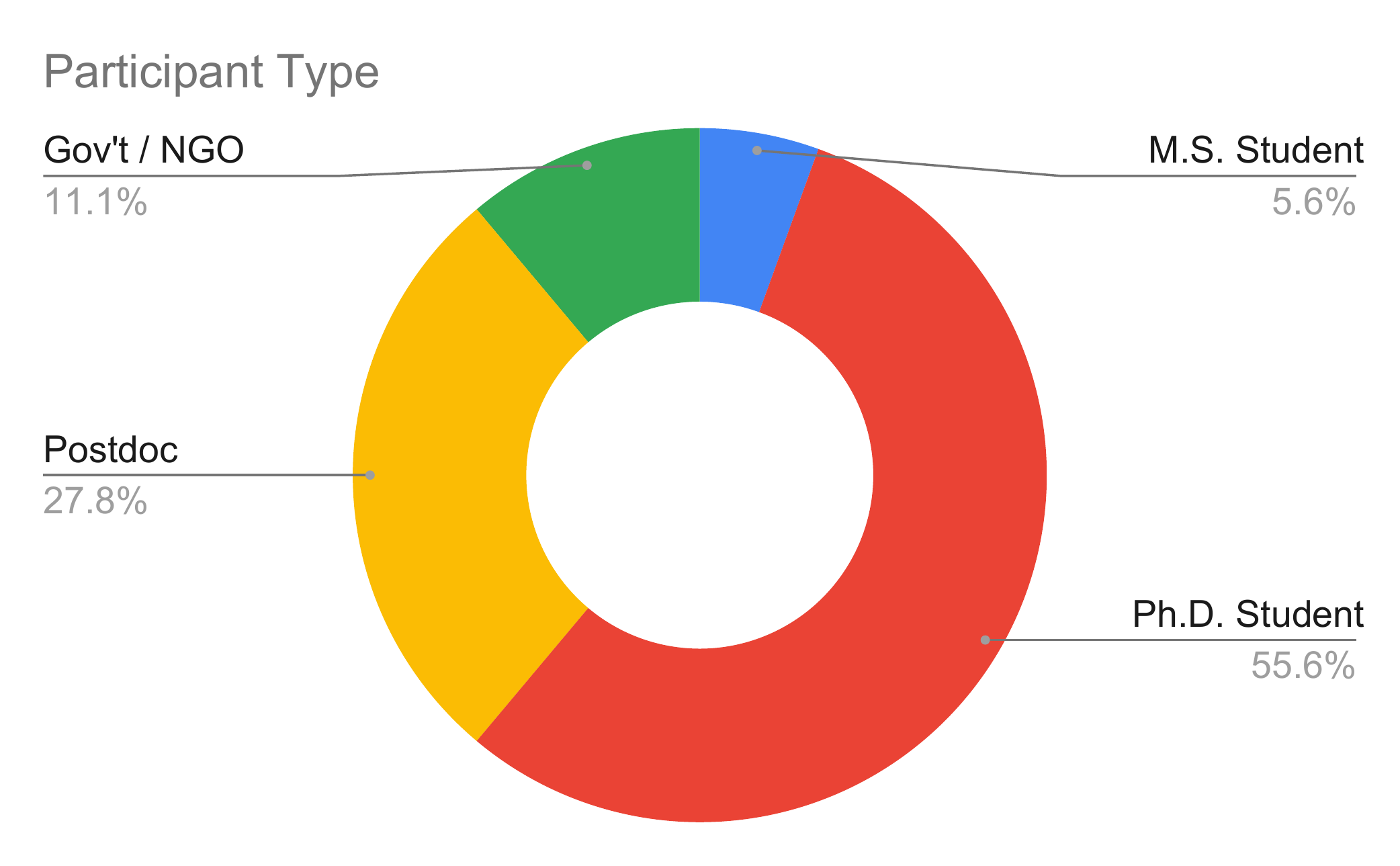}
\includegraphics[width=0.49\textwidth]{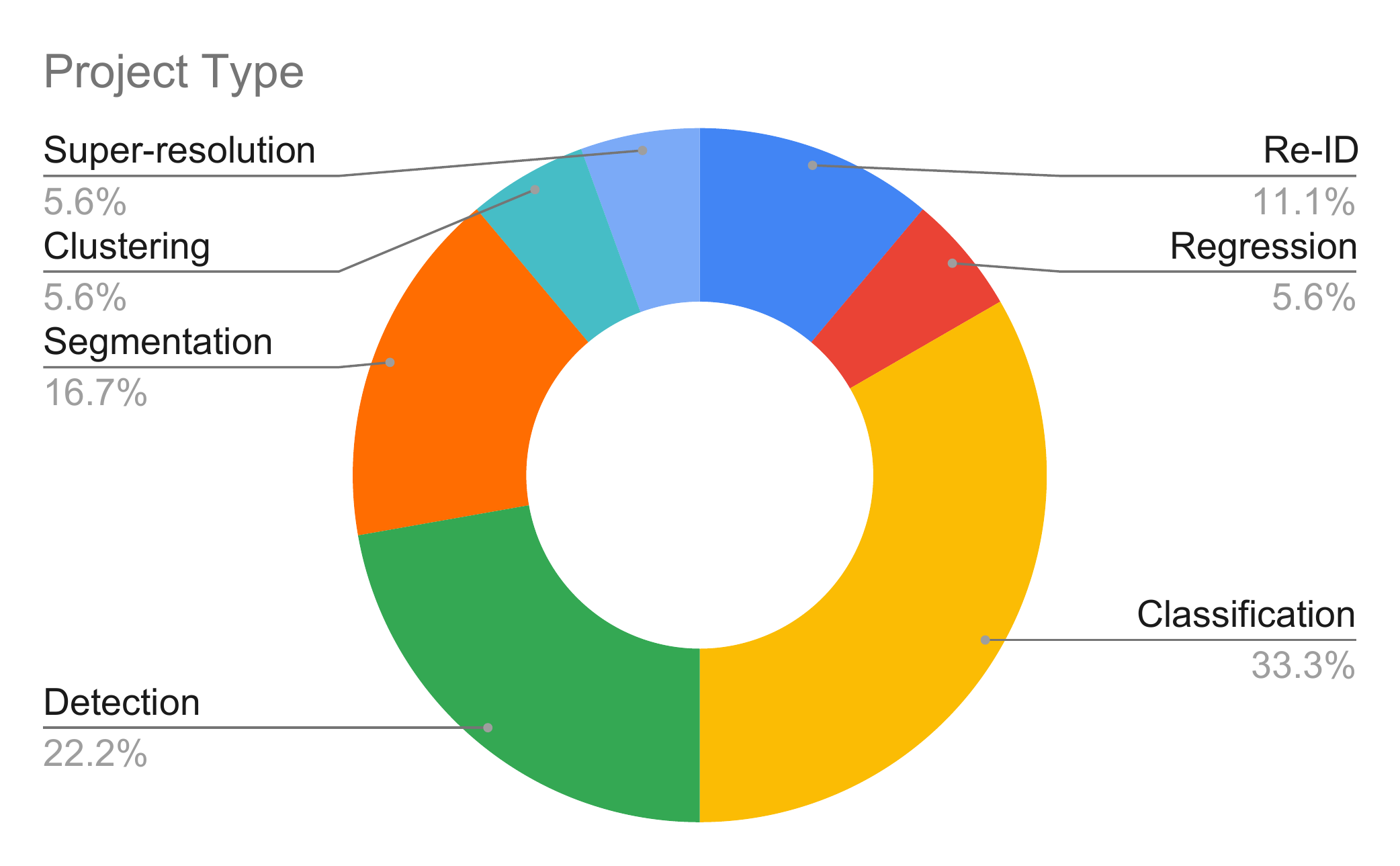}
\caption{Summary of the 2022 CV4E Workshop participant backgrounds (top) and project categories (bottom). Full details can be found in Appendix~\ref{appendix:cohort}.}
\label{fig:cohort}
\end{figure}

\textbf{Selection process.} The CV4E staff recruited application reviewers from the machine learning and ecology communities. Each application received two reviews. Final decisions were made by the CV4E staff. The primary criteria were: (i) goal clarity, (ii) project feasibility, (iii) potential impact, and (iv) candidate preparation. Details about the 2022 cohort can be found in Figure~\ref{fig:cohort} and Appendix~\ref{appendix:cohort}. To maximize accessibility, all participants were funded for travel, room, and board for the duration of the program.

\textbf{Pre-workshop preparation.} All participants were added to a Slack workspace which served as the primary communication channel for the workshop. Each participant was assigned to a working group overseen by a CV4E instructor. During the $\sim$6 months between participant selection and the beginning of the workshop, participants met with their instructors to finalize project plans and address any data or label issues. Participants were also expected to learn Python during this period. Instructors assisted by providing Python resources and holding biweekly office hours. 

\begin{figure}[htb]
\centering
\includegraphics[trim={6.5cm 13cm 6.5cm 1.8cm},clip,width=0.49\textwidth]{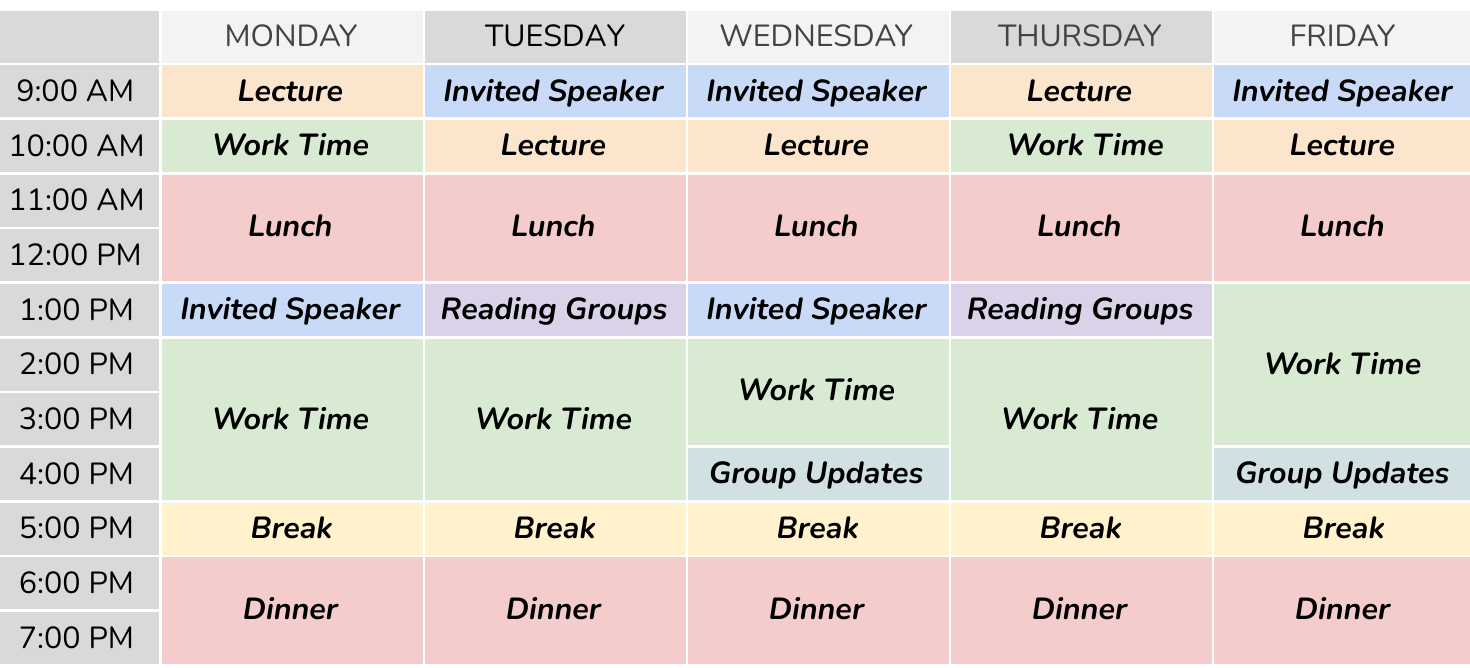}
\caption{The weekly schedule for the 2022 CV4E Workshop was roughly evenly split between instructional time (\emph{Lectures}, \emph{Invited Speakers}, \emph{Reading Groups}, \emph{Group Updates}) and instructor-supervised working time (\emph{Work Time}). In practice, portions of the \emph{Group Updates}, \emph{Lunch}, \emph{Break}, and \emph{Dinner} slots were often used by participants as extra work time.}
\label{fig:schedule}
\end{figure}

\textbf{In-person workshop.} Figure~\ref{fig:schedule} gives a representative weekly schedule for the CV4E Workshop. Participants received classroom instruction from \emph{Lectures} and \emph{Invited Speakers}. Each participant joined a \emph{Reading Group} on a topic of their choice (see Appendix~\ref{appendix:reading}), which met twice weekly for a guided discussion of research papers. During the \emph{Work Time}, participants worked on their projects independently, with CV4E staff and working groups peers available for questions. Each working group discussed their progress and obstacles during the \emph{Group Updates}. 

\textbf{Outcomes.} All 18 of our participants had trained models for their projects by the end of the workshop. Some of these models were already achieving high performance, while others needed more investigation. In addition, the participants and staff formed a community that has endured beyond the workshop through the Slack workspace and ongoing projects. 

\section{Lessons Learned}\label{sec:lessons_learned}

\textbf{Enforce structured Python preparation.} The primary obstacle for most participants was insufficient Python preparation. While participants were not required to know Python before applying, they were asked to learn Python before arriving. To facilitate this process, the staff provided resources for learning Python and hosted office hours in the months leading up to the CV4E Workshop. However, many participants (even capable R programmers) still struggled with Python issues throughout the workshop. In hindsight, we overestimated the extent to which R experience is helpful for quickly learning Python. In the future we will enforce more structured Python preparation before the workshop. 

\textbf{Start simple.} It is challenging to build a working computer vision system from scratch in 3 weeks. To maximize the probability of success, it is important to start simple. When appropriate, we encouraged participants to use standard well-understood pipelines \eg fine-tuning an ImageNet-pretrained ResNet-50. 

\textbf{Work in long blocks.} Participants made much more progress during long blocks of work time (3$+$ hours) than during shorter work blocks (1-2 hours).

\textbf{Collect similar projects in working groups.} Participants were often eager to help each other, especially when they were deploying similar techniques. Working groups should be constructed to maximize opportunities for such collaborations. 

\textbf{Mix experience levels in working groups.} Some participants had significant experience with machine learning or programming, enabling them to make swift progress on their projects with minimal assistance from instructors. Experienced participants routinely volunteered to assist less experienced participants, which seemed mutually beneficial. In the future, we plan to ensure that each working group has a mix of experienced and inexperienced participants. 

\textbf{Make unambiguous infrastructure recommendations.} There are many reasonable ways to set up the infrastructure necessary for computer vision work. For instance, consider the problem of developing code which is meant to be executed on a VM. One approach is to edit the code locally in a text editor and move it to the VM using \texttt{rsync}, handling revision control locally. Another approach is to use a tool like VSCode \cite{VSC_web} which allows code on the VM to be edited directly via SSH. In this case, revision control would be handled on the VM. A third approach is to edit code locally, \texttt{push} the code to GitHub, and \texttt{pull} the code to the VM. Revision control is ``built in" for this workflow. The instructors had different preferences, and no workflow was clearly superior. Participants did not benefit from being asked to make their own choice about which workflow to use. In the future we will provide unambiguous and unified recommendations for development infrastructure. 

\textbf{Avoid deep learning library wrappers.} There are many ``wrappers" for deep learning libraries which are meant to make deep learning tools easier to use. Some are general-purpose (\eg PyTorch Lightning~\cite{PTL_web}) while others are domain-specific (\eg  OpenSoundscape~\cite{OSS_web}, DeepForest~\cite{DF_web}, TorchGeo~\cite{TG_web}). While these wrappers are undoubtedly useful, they are not ideal for our participants for two reasons. First, they conceal too much complexity which hinders the process of learning about \eg training loops and data flow. Second, they were more difficult to customize and debug, even with instructor assistance. In the future, we will encourage all participants to work directly with deep learning libraries.

\textbf{Avoid Jupyter Notebooks.} Jupyter Notebooks provide capabilities familiar to experienced R users, such as the ability to run sections of code interactively. However, participants who relied on Jupyter Notebooks while learning Python often struggled to transition to more traditional command line workflows when developing their computer vision systems. We now believe that learning to work with Python through the command line provides a better foundation for understanding machine learning workflows.

\textbf{Make sure GPUs are available.} Cloud computing services like AWS and Azure often provide free credits for education and research. However, GPUs may not be available depending on customer demand. It is important to confirm with cloud providers that GPUs will be made available. Alternatively, consider using local computing resources or university clusters.

\section{Educational Techniques}\label{sec:educational_techniques}

In this section we describe a few educational techniques we found helpful for the CV4E Workshop.

\textbf{Guided troubleshooting.} Troubleshooting and debugging are vital skills in machine learning, and it was important to provide participants with opportunities to hone these abilities. However, due to the tight schedule of the CV4E Workshop, we did not want participants to be stuck on any one problem for too long. To balance these objectives, instructors tried to walk participants through the troubleshooting process by asking leading questions about the problems they were encountering. For unusual problems of limited educational value (\ie complex configuration or installation issues), instructors intervened to resolve the issue as quickly as possible.

\textbf{Pair pseudocoding.} Most of our participants were not comfortable writing Python code at the beginning of the CV4E Workshop, so we wanted to provide frequent opportunities for hands-on coding. Whenever possible, instructors avoided writing code for the participants. To prevent participants from getting stuck on code design issues, we used \emph{pair pseudocoding}:
\begin{enumerate}
    \item The instructor asks the participant to explain what they would like to accomplish,  discussing until the goal is clear to both parties. 
    \item The instructor writes pseudocode that solves the problem and walks through it with the participant to help them understand the logic of the solution. The pseudocode can be more specific or vague depending on the participant's needs.
    \item The participant writes Python code to solve the problem, while the instructor remains available to answer questions as they arise. 
\end{enumerate}

\textbf{Goal statements.} During the initial stages of the project, some of our participants felt like progress was not being made because the code didn't ``work" yet. To make their progress more salient, some instructors asked participants to make a \emph{goal statement} at the beginning of each work session, and to check progress towards that goal at the end of each work session. This strategy helped participants to maintain motivation until more tangible results were obtained. 

\textbf{Contextualized lectures.} Maintaining interest during lectures was not a significant problem for the CV4E Workshop due to the enthusiasm of the participants. However, it is easy for lectures on machine learning topics to become too abstract. We tried to ensure that the lectures remained grounded in applications and examples. Since each participant had their own applied problem in mind, we often paused lectures to ask participants to reflect on how the lecture topic applied to their individual projects. Participants shared their answers with the class, providing concrete examples that illustrated the lecture topic. 

\section{Conclusion}\label{sec:conclusion}
We have described our experience at the inaugural Resnick Sustainability Institute Summer Workshop on Computer Vision Methods for Ecology. We consider the format to be a success, as all of our participants trained models for their projects by the end of the workshop. However, we have also discussed some challenges we encountered and identified opportunities to improve the CV4E Workshop. We hope these observations will be useful for others who teach computer vision across disciplines. 

\section{Acknowledgements}

We would like to thank the Resnick Sustainability Institute, Caroline Murphy, Xenia Amashukeli, and Pietro Perona for making the CV4E Workshop possible. Computing credits were provided by Amazon AWS and Microsoft Azure. We also thank the inaugural cohort of the CV4E Workshop: Ant\'on \'Alvarez, Carly Batist, Peggy Bevan, Catherine Breen, Anna Boser, Tiziana Gelmi Candusso, Melanie Clapham, Rowan Converse, Roni Goldshmid, Natalie Imirzian, Brian Lee, Francesca Ponce, Alixandra Prybyla, Rachel Renne, Felix Rustemeyer, Taiki Sakai, Ethan Shafron, and Casey Youngflesh. 

{
\raggedright
\bibliographystyle{plain}
\bibliography{main}
}

\clearpage
\appendix

\section{2022 Cohort}\label{appendix:cohort}

\subsection{Participant Backgrounds}
The inaugural 2022 CV4E Workshop had 18 participants. Broken down by current occupation, our cohort consisted of: 
\begin{itemize}
    \item 1 Master's student; 
    \item 10 Ph.D. students; 
    \item 5 post-doctoral researchers; and 
    \item 2 researchers from government agencies or non-governmental organizations.
\end{itemize}
Broken down geographically, our cohort consisted of:
\begin{itemize}
    \item 11 participants from 7 different states in the U.S.; 
    \item 5 participants from European countries; and 
    \item 2 participants from Canada.
\end{itemize}
Participants came from diverse academic backgrounds, including conservation biology, biological anthropology, geography, mechanical engineering, civil engineering, neuroscience, and ecology.

\subsection{Participant Projects}
Projects fell into seven main categories. 
\begin{enumerate}
\item {\bf Individual Re-Identification:} Associating images of the same animal taken from various cameras, locations, and times. The two relevant projects were: (1) re-identifying bears, (2) re-identifying Iberian Lynx. 
\item {\bf Regression:} Assigning a continuous number to an image or video. The one relevant project was: (1) analyzing wind speed from overhead drone video of trees. 
\item {\bf Classification:} Categorizing or labeling images or parts of images from a fixed collection of categories. The six relevant projects were: (1) determining presence or absence of lemur vocalizations, (2) beaked whale species classification from echolocation clicks, (3) bumblebee species and caste classification from flight sounds, (4) assigning ants to size categories, (5) identifying weather conditions from camera trap images, and (6) species identification in urban camera traps. Note that projects (1), (2), and (3) used computer vision techniques to classify images (spectrograms) that represent audio signals. 
\item {\bf Object Detection:} Locating instances of objects in images or videos. The four relevant projects were detecting: (1) piospheres, (2) woodland draws, (3) flies, and (4) waterfowl. Projects (3) and (4) use detection as an intermediate step towards counting.  
\item {\bf Segmentation:} Classifying pixels based on their semantic characteristics. The three relevant projects were segmenting: (1) walrus groups in the Arctic, (2) permafrost, and (3) trees. All projects were based on remote sensing imagery. 
\item {\bf Clustering:} Grouping objects together according to some notion of similarity. The one relevant project was: (1) determining the species richness of an area using the number of clusters in a collection of camera trap imagery. 
\item {\bf Super-resolution:} Increasing the resolution of an image. The one relevant project was: (1) increasing the resolution of land surface temperature data using satellite imagery.
\end{enumerate}

\section{Key Topics}\label{appendix:key_skills}
In this section we catalog topics that many of our participants learned during the workshop, either through formal instruction or on their own. We emphasize tools and concepts that were initially unfamiliar to most participants. For each topic, we describe the content and explain why it was important for our participants. See also the list of lectures in Appendix~\ref{appendix:lectures}.

\subsection{Tools}

\subsubsection{Annotation Tools}

\noindent 
\textbf{Content}: Using annotation tools to label image~\cite{imglab_web, cvat_web} or audio~\cite{audacity_web} data. 

\noindent
\textbf{Motivation}: Labeled data is essential for training and evaluating computer vision algorithms. Since CV4E Workshop participants were using their own data, many of them needed to learn to use some sort of annotation tool. Furthermore, many of these tools can export labels in the standard formats expected by open-source computer vision libraries. 

\subsubsection{Unix Commands}

\noindent
\textbf{Content}: Common Unix commands like \texttt{ls}, \texttt{pwd}, \texttt{rm}, \texttt{mkdir}, \texttt{rmdir}, \texttt{mv}, \texttt{cat}, \texttt{head}, \texttt{tail} etc. Occasionally, less common commands like \texttt{chmod} or \texttt{grep}. 

\noindent
\textbf{Motivation}: Facility with Unix commands is crucial for installing packages, working with virtual machines, and using revision control. Understanding Unix commands also helps to build intuition for core concepts like absolute vs. relative paths. 

\subsubsection{Terminal-Based Text Editing}

\noindent
\textbf{Content:} Tools like \texttt{nano} for editing text that is stored on a server from the command line. 

\noindent
\textbf{Motivation:} When configuring SSH authentication it is often necessary to edit text files on the VM (\eg  $\sim$\verb|/.ssh/config|). 

\subsubsection{Terminal Multiplexing}

\noindent
\textbf{Content:} Tools like \texttt{tmux} or \texttt{screen} for managing terminal sessions.

\noindent
\textbf{Motivation:} Long-running code (\eg model training in PyTorch) should be executed in a terminal session that is decoupled from the SSH connection to avoid being terminated when a laptop is closed or internet connection is lost. 

\subsubsection{SSH}

\noindent
\textbf{Content:} The \texttt{ssh} command and SSH keys. Occasionally, SSH tunneling. 

\noindent
\textbf{Motivation:} The \texttt{ssh} command is used to create a terminal session connected to a VM. Related topics like SSH keys are also important for \eg authenticating terminal-based file transfers and enabling GitHub access. SSH tunneling can be necessary for setting up tools like TensorBoard \cite{TB_web}. 

\subsubsection{Terminal-Based File Transfers}

\noindent
\textbf{Content:} Tools like \texttt{scp} or \texttt{rsync} for transferring files. 

\noindent
\textbf{Motivation:} Command line tools are the most reliable way to move large amounts of data from one place to another. This is useful for local transfers (\eg from one hard drive to another) and remote transfers (\eg from a local hard drive to a storage volume attached to a virtual machine). 

\subsubsection{Revision Control}

\noindent
\textbf{Content:} Using GitHub for tracking changes made to code. 

\noindent
\textbf{Motivation:} Code for computer vision projects tends to quickly grow in complexity, and it is easy to forget what has changed since the last working version. Tools like GitHub allow earlier versions of the code to be revisited easily if a bug was introduced by some change. In addition, GitHub can be used to move code from a local machine (\texttt{git push}) to a virtual machine (\texttt{git pull}) along with allowing users to download (\texttt{git clone}) open-sourced computer vision repositories. 

\subsubsection{Cloud Computing}

\noindent
\textbf{Content:} Interacting with the web interfaces of cloud computing providers. Creating a virtual machine with appropriate resources \eg GPUs, storage. Estimating and managing cost. 

\noindent
\textbf{Motivation:} One of the most common ways to access GPU resources for computer vision work is to use a VM from a cloud computing provider like Amazon Web Services (AWS) \cite{AWS_web} or Microsoft Azure \cite{AZR_web}. It is important to understand the benefits (scalability, reliability) and drawbacks (cost) of cloud computing.

\subsubsection{Virtual Environments}

\noindent
\textbf{Content:} Creating and managing virtual environments. 

\noindent
\textbf{Motivation:} Computer vision projects typically rely on large pre-existing codebases, which may require particular versions of certain packages to be installed. While the user could change their base installations, a better solution is to create a virtual environment (through \eg \texttt{conda}) in which the dependencies of the codebase can be installed. Virtual environments are also useful if a ``clean reinstall" becomes necessary, because they are easy to create and delete. 

\subsubsection{Python}

\noindent
\textbf{Content:} Basic syntax, conditionals, loops, string parsing, file I/O, functions, classes and data structures.

\noindent
\textbf{Motivation:} Facility with Python is crucial for efficiently working with Python-based deep learning libraries, which the computer vision community uses almost exclusively. 

\subsubsection{Python Libraries}

\noindent
\textbf{Content:} Common libraries like \texttt{numpy}, \texttt{pandas}, \texttt{ipdb}, \texttt{sklearn}, and \texttt{matplotlib}. 

\noindent
\textbf{Motivation:} Python has many stable, high-quality libraries for numerical computing and data analysis. Libraries like \texttt{ipdb} allow for in-line debugging. 

\subsubsection{Deep Learning Libraries}

\noindent
\textbf{Content:} Preferably PyTorch \cite{PT_web} and alternatively TensorFlow \cite{TF_web} for building deep learning systems. 

\noindent
\textbf{Motivation:} Modern deep learning libraries are indispensable for developing and training computer vision systems.

\subsubsection{Image Processing Libraries}

\noindent
\textbf{Content:} Libraries and command-line tools like OpenCV~\cite{openCV_web}, ImageMagick~\cite{IM_web}, and FFmpeg~\cite{ffmpeg_web}. 

\noindent
\textbf{Motivation:} These tools are often used for efficient data augmentation and visualization. 

\subsection{Computer Science Concepts}

There are a few core concepts from computer science that came up frequently throughout the program. 

\subsubsection{Object Oriented Programming}

\noindent
\textbf{Content:} Classes and  objects. Inheritance, encapsulation, polymorphism. 

\noindent
\textbf{Motivation:} Many important libraries assume an understanding of object oriented programming concepts. For instance, one common point of confusion for our participants was the difference between the PyTorch dataset class and a dataset object from that class. Understanding object oriented programming also makes it easier to understand data structures. 

\subsubsection{Data Structures}

\noindent
\textbf{Content:} Common data structures (\eg list, tuple, dictionary, NumPy array, PyTorch tensor) and their methods, casting from one data type to another, checking data structures. 

\noindent
\textbf{Motivation:} Unexpected behavior differences between \eg Python lists, NumPy arrays, and PyTorch tensors can cause significant frustration if data structures are not well understood. 

\subsubsection{Data Types}

\noindent
\textbf{Content:} Common data types \eg int, float, double, string, and bool.

\noindent
\textbf{Motivation:} Understanding data types increases context understanding and can significantly impact data storage size.

\subsubsection{Namespaces}

\noindent
\textbf{Content:} The built-in, global, and local namespaces. 

\noindent
\textbf{Motivation:} Namespaces are the answer to many common questions \eg why variables defined inside a function are not accessible outside the function. 

\subsubsection{Mutability}

\noindent
\textbf{Content:} Mutable and immutable objects. In-place operations. 

\noindent
\textbf{Motivation:} Mutable objects can be changed in-place while mutable objects cannot. This is the basis for understanding whether changes made to an object inside a function will affect the object outside of the function. 

\subsection{Machine Learning Concepts}

Participants learned different practical and conceptual aspects of computer vision and machine learning depending on their project. However, all participants had to engage with a few core concepts. 

\subsubsection{Generalization}

\noindent
\textbf{Content:} The concept of generalization, different types of generalization, identifying a type of generalization that reflects the goals of a project.

\noindent
\textbf{Motivation:} In ecology there are many different notions of generalization, and it is important to choose one that reflects the goals of a project. For instance, in camera trap image classification it might be important to generalize to new locations or to future data from the same locations. These different notions of generalization need to be measured in different ways. 

\subsubsection{Data Splits}

\noindent
\textbf{Content:} The role of training, validation, and testing data. Designing appropriate splits to measure the chosen type of generalization. 

\noindent
\textbf{Motivation:} Training, validation, and testing splits should be designed to capture an appropriate problem-specific notion of generalization. These splits must then be handled appropriately (\eg no hyperparameter tuning on the test split) to ensure that performance measurements reflect generalization. 

\subsubsection{Overfitting}

\noindent
\textbf{Content:} Defining and recognizing overfitting. Mitigating overfitting using regularization techniques. 

\noindent
\textbf{Motivation:} All participants were working with deep learning, for which overfitting is always a significant concern. 

\subsubsection{Evaluation Metrics}

\noindent
\textbf{Content:} Common evaluation metrics for different tasks, their strengths and limitations, choosing metrics that reflect high-level goals. 

\noindent
\textbf{Motivation:} Appropriate metrics are vital for determining which approaches work best and deciding if a computer vision system is ``good enough" to be used for a real application. 

\subsubsection{Deep Learning}

\noindent
\textbf{Content:} Neural networks, loss functions, minibatch gradient descent. 

\noindent
\textbf{Motivation:} All modern computer vision methods rely on deep learning. Since our participants were building and troubleshooting computer vision systems, they needed to understand deep learning basics as well. Loss functions were a particular focus, since changing the loss is one of the primary ways of adapting an existing method to a new problem. 

\subsubsection{Representations}

\noindent
\textbf{Content:} Image embeddings, distances in embedding space, pretraining, transfer learning. 

\noindent
\textbf{Motivation:} ImageNet pretraining is ubiquitous in modern computer vision, but many of our participants work in specialized domains for which ImageNet pretraining may not be appropriate. Domain-specific pretraining requires an understanding of representation learning. The concept of image embeddings is also useful for understanding many common computer vision algorithms (\eg metric learning) and visualization techniques (\eg t-SNE). 

\subsection{Other Skills}

\subsubsection{Critically Reading Machine Learning Papers}

\noindent
\textbf{Content:} Understanding machine learning terminology and paper structure, critically interpreting claims, evaluating complexity vs. performance trade-offs.

\noindent
\textbf{Motivation:} Exploring the literature in a new field is always daunting. This is particularly challenging in machine learning where papers may be over-enthusiastically written, necessitating extra vigilance from the reader to clearly understand the drawbacks and benefits of a method~\cite{lipton2018troubling}. 

\subsubsection{Selecting ``Good" Open Source Libraries}

\noindent
\textbf{Content:} Recognizing markers of quality in open source code.

\textbf{Motivation:} There is plenty of open-source computer vision code, but not all of it is reliable or well-maintained. Participants must learn to check indicators of code quality \eg how many users a library has or how often the developers fix bugs.

\subsubsection{Digging in to Libraries}

\noindent
\textbf{Content:} Reading documentation, finding the code that handles a certain task, understanding how components of a codebase interact. 

\noindent
\textbf{Motivation:} Computer vision projects depend on numerous complex but (generally) well-documented libraries. It is important to be able to understand the documentation. Sometimes it also becomes necessary to locate and inspect the piece of code being documented (\eg a function from some library) to understand how it works in detail.

\subsubsection{General Troubleshooting}

\noindent 
\textbf{Content:} Errors vs. warnings, searching for more information about error messages. 

\noindent
\textbf{Motivation:} Errors and warnings are common when \eg installing packages or testing new code. One of the most important skills in any programming activity is the ability to use a search engine to understand an error message. This involves locating the appropriate part of an error message to use as a search term, reading through the results, and choosing an appropriate next step. 

\subsubsection{Debugging Python Code}

\noindent
\textbf{Content:} Types of errors, finding the source of an error, print statement debugging. 

\noindent
\textbf{Motivation:} For a given line of code, any number of errors could arise. Understanding the different types of Python errors is helpful for pinpointing the root cause. Print statement debugging is also extremely useful for troubleshooting code running on a remote machine. 

\section{List of Lectures}\label{appendix:lectures}

\begin{enumerate}
    \item Intro and Logistics (Sara Beery)
    \item Dataset Prototyping and Visualization (Jason Parham)
    \item Working on the Cloud (Suzanne Stathatos)
    \item Data Splitting and Avoiding Data Poisoning (Sara Beery)
    \item Training your Model: Deciding on Configurations, Launching, Monitoring, Checkpointing, and Keeping Runs Organized (Benjamin Kellenberger)
    \item Working with Open-Source CV Codebases: Choosing a Baseline Model and Custom Data Loading (Sara Beery)
    \item Evaluation Metrics (Elijah Cole)
    \item Offline Evaluation and Analysis (Sara Beery)
    \item What's next? Rules of Thumb to Improve Results (Benjamin Kellenberger)
    \item Data Augmentation (Bj\"orn L\"utjens)
    \item Expanding and Improving Training Datasets with Models: Weak Supervision, Self Supervision, Targeted Relabeling, and Anomaly Detection (Tarun Sharma)
    \item Fair Comparisons and Ablation Studies: Understanding What is Important (Elijah Cole)
    \item Efficient Models: Speed vs. Accuracy (Justin Kay)
    \item Serving, Hosting, and Deploying Models and Quality Control (Jason Parham)
\end{enumerate}

\section{List of Reading Groups}\label{appendix:reading}

\begin{enumerate}
    \item Time Series, Spectral Transforms, and Remote Sensing
    \item Data Imbalance \& Long Tail Distributions
    \item Weak Supervision, Unsupervised Learning, Fine-tuning \& Transfer Learning
    \item Bias \& Domain Shift and Generalization
\end{enumerate}

\end{document}